\documentclass[aps,prl,reprint,amsmath,amssymb,superscriptaddress,nobibnotes,showpacs]{revtex4-1}

\usepackage{graphicx}
\usepackage{dsfont}    
\usepackage{color}

\begin{document}

\title{Observation of Young's Double-Slit Interference with the Three-Photon N00N State}

\author{Yong-Su Kim} 
\affiliation{Department of Physics, Pohang University of Science and Technology (POSTECH), Pohang, 790-784, Korea } 

\author{Osung Kwon}
\affiliation{Department of Physics, Pohang University of Science and Technology (POSTECH), Pohang, 790-784, Korea }

\author{Sang Min Lee}
\affiliation{Division of Convergence Technology, Korea Research Institute of Standards and Science, Daejeon, 305-340, Korea }

\author{Heonoh Kim}
\altaffiliation[Present address: ]{Department of Physics, University of Ulsan, Ulsan, 680-749, Korea}
\affiliation{Division of Convergence Technology, Korea Research Institute of Standards and Science, Daejeon, 305-340, Korea }

\author{Sang-Kyung Choi}
\affiliation{Division of Convergence Technology, Korea Research Institute of Standards and Science, Daejeon, 305-340, Korea }

\author{Hee Su Park}\email{hspark@kriss.re.kr}
\affiliation{Division of Convergence Technology, Korea Research Institute of Standards and Science, Daejeon, 305-340, Korea }

\author{Yoon-Ho Kim}\email{yoonho72@gmail.com}
\affiliation{Department of Physics, Pohang University of Science and Technology (POSTECH), Pohang, 790-784, Korea }

\date{\today}

\begin{abstract}
Spatial interference of quantum mechanical particles exhibits a fundamental feature of quantum mechanics. A two-mode entangled state of N particles known as N00N state can give rise to non-classical interference. We report the first experimental observation of a three-photon N00N state exhibiting Young's double-slit type spatial quantum interference. Compared to a single-photon state, the three-photon entangled state generates interference fringes that are three times denser. Moreover, its interference visibility of $0.49 \pm 0.09$ is well above the limit of 0.1 for spatial super-resolution of classical origin. The demonstration of spatial quantum interference by a N00N state composed of more than two photons represents an important step towards applying quantum entanglement to technologies such as lithography and imaging.
\end{abstract}



\maketitle


Double-slit interference exhibited by single-photons or single-electrons is one of the most fundamental effects in quantum physics and is intimately tied to many foundational concepts in quantum physics such as complementarity, the uncertainty principle, and Born's rule \cite{feynman,kim00,sinha10}. The number-path entangled state or the N00N state $|\psi\rangle = \frac{1}{\sqrt{2}}(|N\rangle_a|0\rangle_b + |0\rangle_a|N\rangle_b)$, where $N$ is the number of quanta and the subscript refers to the spatial mode, naturally arises in generalizing the double-slit experiment to the $N$-quantum case and was first discussed in the context of photonic de Broglie waves \cite{jacobson95}. Such states are of fundamental importance in quantum physics as they represent macroscopic quantum superposition or Schr\"odinger `cat' states \cite{ourj}. Moreover, recent research has shown that the N00N state is at the heart of many quantum-enhanced measurement schemes \cite{dow}. For instance, quantum lithography which enables drawing of arbitrary high-contrast patterns with resolution beyond the classical Rayleigh limit requires the N00N state \cite{boto00,bjork01}. Also, various quantum metrology schemes aimed at achieving the Heisenberg-limited sensitivity are based on the use of the N00N state \cite{giovannetti04,kwon10}.

Many ideas have been proposed on how to prepare the photonic N00N state \cite{dow,fiu,cable07,kapale07,dan}. Experimental demonstrations of the N00N state, however, have been rather limited. To date, up to five-photon N00N states have been reported in literature \cite{eda02,dangelo01,kawabe07,mitchell04,kim09, walther04, sun06,nagata07, afek10}. However, for four- and five-photon N00N states, experimental demonstrations so far are limited to the measurement-based projection of the N00N state \cite{walther04, sun06, nagata07, afek10}.



The Young-type double-slit experiment demonstrating spatial quantum interference of the N00N state as originally proposed for quantum lithography, on the other hand, has only been reported for the $N=2$ N00N state using spontaneous parametric down-conversion (SPDC) \cite{dangelo01,kawabe07}. Despite scientific importance, the $N=2$ case based on SPDC does not offer any resolution breakthrough because it simply retrieves the resolution already obtainable with the pump laser. For truly demonstrating quantum-enhancement of spatial resolution beyond the classical limit, it is essential to show spatial interference fringes for the N00N state with $N \ge 3$. However, all known N00N-related experiments for $N>2$ reported to date have dealt exclusively with Mach-Zehnder type interferometers with a phase shifter fixed in space, thus only exhibiting temporal interference fringes.


In this paper, we report the first experimental demonstration of the Young's double-slit type spatial quantum interference of the three-photon N00N state, exhibiting three times denser spatial interference fringes than that of the single-photon state.  This is the first time, to the best of our knowledge, that the spatial quantum interference of the N00N state is observed for more than two photons, thus paving the way towards quantum optical interferometric lithography and quantum imaging with the N00N state. 

\begin{figure}[t]
\includegraphics[width=3in]{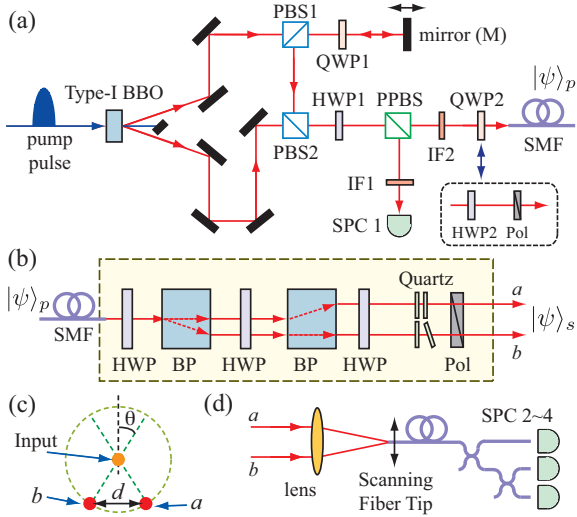}
\caption{Schematic of the experimental setup. (a) The polarization N00N state $|\psi\rangle_p$ is prepared in a single spatial mode. IF1 (IF2) is an interference filter with full-width at half-maximum bandwidth of 5 nm (10 nm) centered at 780 nm. (b) $|\psi\rangle_p$ is losslessly transformed to $|\psi\rangle_s$ using a mode converter (top view). (c) Front view of the mode converter. (d) The Young-type double-slit interference is observed at the focus of a lens. }\label{setup}
\end{figure}

The schematic of the experimental setup is shown in Fig.~\ref{setup}. First, two pairs of horizontally polarized photons centered at 780 nm are generated at the 2 mm thick type-I BBO crystal via the SPDC process pumped by a femtosecond pump pulse centered at 390 nm, see Fig.~\ref{setup}(a). The pump pulse has a duration of 200 fs and a repetition rate of 76 MHz. The two pairs of SPDC photons are then brought back together in a single spatial mode with the help of polarizing beam splitters (PBS1 and PBS2), a quarter-wave plate (QWP1) oriented at $45^\circ$, and a movable mirror (M). The half-wave plate HWP1 is oriented at $22.5^\circ$ and the partially-polarizing beam splitter (PPBS) has an amplitude reflection coefficient of $\sqrt{2/3}$ for vertical polarization. If the mirror M is set so that the delay time between the two pairs of SPDC photons are zero, the photon state before HWP1, $|\psi\rangle_i = \frac{1}{2} a_H^{\dag 2} a_V^{\dag 2} |0\rangle$, is transformed by passing through the PPBS to become
\begin{eqnarray}
|\psi\rangle &= &\frac{\sqrt{2}}{6}(a_{V1}^\dag  (\frac{1}{3} a_{V2}^{\dag 3} - a_{H2}^{\dag 2} a_{V2}^{\dag}) ) |0\rangle\nonumber\\
& & + (\frac{1}{8}a_{H2}^{\dag 4} -\frac{1}{12}a_{H2}^{\dag 2} a_{V2}^{\dag 2}+\frac{1}{72}a_{V2}^{\dag 4} )|0\rangle + |\textrm{etc}\rangle,
\label{state}
\end{eqnarray}
where $a_H^\dag$ ($a_V^\dag$) refers to the creation operator for a horizontal (vertical) polarized photon, $|0\rangle$ is the vacuum, and the subscripts 1 (2) refers to the reflected  (transmitted) mode of the PPBS \cite{kim09}. Note that, in eq.~(\ref{state}), the four-photon amplitudes that do not result at least three photons in mode 2 are expressed as $|\textrm{etc}\rangle$ and they do not contribute to the $N=3$ N00N state interference as they cannot be registered at the three-photon detector.

The first term in eq.~(\ref{state}) is now relevant to our purpose: if a single-photon detection occurs at the single-photon counter SPC1, the three-photon state in mode 2 after the quarter-wave plate QWP2 oriented at $45^\circ$ is given as  
\begin{equation}
|\psi\rangle_p = (|3\rangle_H |0\rangle_V + i|0\rangle_H |3\rangle_V)/\sqrt{2}.\label{pN00N}
\end{equation}
In other words, the three-photon polarization N00N state $|\psi\rangle_p$ is heralded in mode 2 (at the entrance of the single mode fiber SMF) whenever there is a single-photon detection at SPC1 \cite{kim09}. A three-photon absorber located at mode 2, if triggered by SPC1, would be able to record genuine three-photon quantum interference fringes due to the N00N state. The second term in eq.~(\ref{state}) is irrelevant to the heralded N00N state preparation but it would inevitably contribute as background noise to the three-photon absorber. (In principle, this can be prevented by installing a trigger-operated shutter in mode 2.)  

Finally, the polarization N00N state $|\psi\rangle_p$ is transformed to the spatial two-mode three-photon N00N state $|\psi\rangle_s$,
\begin{equation}
|\psi\rangle_s = (|3\rangle_a |0\rangle_b + |0\rangle_a |3\rangle_b)/\sqrt{2},\label{sN00N}
\end{equation}
by using the mode converter shown in Fig.~\ref{setup}(b) and Fig.~\ref{setup}(c). The mode converter, which consists of half-wave plates (HWP), birefringent prisms (BP), quartz plates, and a polarizer (Pol), is based on the following operation principle.  The first and second BPs are aligned so that their optic axes are oriented respectively at $+\theta$ and $-\theta$ with respect to the vertical polarization. The first two HWPs are used for rotating the horizontal-vertical polarization basis of the photons so that they overlap with the rotated optic axes of BPs. The third HWP and the quartz plates are used to match the polarization states of photons in spatial modes $a$ and $b$. An additional horizontal polarizer (Pol) is used to clean up the polarization state so that all three photons in eq.~(\ref{sN00N}) are guaranteed to have the same polarization. 

The spacing $d=2L\sin\theta$ between the modes $a$ and $b$ is determined by the angle $\theta$ and the beam walk-off (i.e., e- and o-ray separation) $L$ of a single BP. In experiment, considering the numerical aperture (0.12) and mode field diameter (5.6 $\mu$m) of SMF, the $1/e^2$ beam diameter of each spatial mode is estimated to be 1.4 mm. The beam spacing $d$ was chosen at $d=2.2$ mm. Our mode converter design provides excellent interferometric phase stability between the spatial modes $a$ and $b$. 
 
To observe the double-slit interference fringes of the three-photon N00N state $|\psi\rangle_s$, a single-mode fiber (identical to SMF) tip was scanned at the focus of a lens (15 mm focal length) by using a piezo-controlled translation stage, see Fig.~\ref{setup}(d). The group delay between the spatial modes $a$ and $b$ was compensated by a set of mirrors (not shown in the figure) in front of the focusing lens. The other end of the fiber tip was connected to three single-photon detectors (SPC2$\sim$4) via a set of 3 dB fiber beamsplitters. The three-fold coincidence SPC2--SPC3--SPC4 triggered by SPC1 constitutes the proper measurement for the heralded three-photon N00N states, $|\psi\rangle_p$ and $|\psi\rangle_s$.


\begin{figure}[t]
   \includegraphics[width=3in]{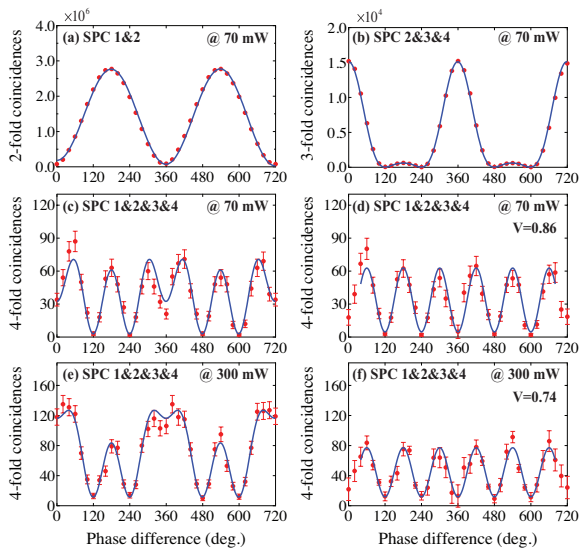}
  \caption{Temporal fringes observed in (a) 2-fold coincidences between SPC1--SPC2, (b) 3-fold coincidences among SPC2--SPC3--SPC4, and (c) $\sim$ (f) 4-fold coincidences. The heralded single-photon state interference is shown in (a). Modulations observed in the 3-fold coincidence shown in (b) come from both the three- and four-photon amplitudes. The heralded three-photon N00N state interference shown in (c) and (e) are plagued with accidental counts due to the four-photon term. The accidental-subtracted three-photon N00N state interference, (d) and (f), show the three-times faster modulation frequency compared to the single-photon interference shown in (a). The data acquisition times (for each data point) are 1200 s for (a) $\sim$ (d) and 100 s for (e) and (f). Solid lines are fitting curves based on eq.~(\ref{prob}). }
 \label{temporal}
\end{figure}

The quality of the polarization N00N state $|\psi\rangle_p$ is directly responsible for the quality of the spatial N00N state $|\psi\rangle_s$, which in turn affects the double-slit interference visibility with $|\psi\rangle_s$. Hence, it is of utmost importance to ensure that the three-photon polarization N00N state is prepared with high purity. Thus, we have first measured the temporal interference fringes due to $|\psi\rangle_p$ as in other N00N state experiments \cite{kwon10,eda02,kim09,mitchell04,walther04,nagata07,sun06,afek10}. 

The first step in preparing the three-photon N00N state is to ensure that photons arrive at PBS2 simultaneously. This can be done by observing the Hong-Ou-Mandel (HOM) interference between detectors SPC1 and SPC2 while scanning the mirror M \cite{kim09,hom}. In the experiment, we observed the HOM dip with 95.4\% visibility at 70 mW pump power.

The heralded three-photon state just before QWP2 is the polarization N00N state in the circular polarization basis, rather than in the linear polarization basis as shown in eq.~(\ref{pN00N}). Thus, to observe temporal quantum interference due to $|\psi\rangle_p$, we replace QWP2 with a half-wave plate (HWP2) and a horizontal polarizer (Pol) as shown in Fig.~\ref{setup}(a) and connect the output end of SMF to the fiber-coupled three-photon detector (SPC2--SPC3--SPC4) depicted in Fig.~\ref{setup}(d). It is then possible to introduce a phase difference $\chi$ between the left-circular and right-circular polarization modes of $|\psi\rangle_p$ by rotating HWP2 by $\chi/4$ and the temporal N00N state interference can be observed in four-fold coincidences between the trigger detector and the three-photon detector.

The experimental data for the temporal interference measurements are shown in Fig.~\ref{temporal}. We first measured, as a reference, the heralded single-photon interference shown in Fig.~\ref{temporal}(a). In Fig.~\ref{temporal}(b), the response of the three-photon detector, three-fold coincidences among SPC2--SPC3--SPC4, is shown. In this case, the first two terms of eq.~(\ref{state}) affect the outcome and the three-fold coincidence probability $P$ is calculated to be 
\begin{eqnarray}
P \propto \eta^3  &&[4 \sin^2\left(3\chi/2\right)+8(\sin\left(\chi\right)+\sin\left(2\chi\right) )^2\nonumber\\
&& +\left(2-\eta\right) (1+2\cos\left(\chi\right) )^4 ],
\label{prob}
\end{eqnarray}
where it is assumed that the three-photon detector consists of three single-photon detectors connected with  3 dB fiber beamsplitters as shown in Fig.~\ref{setup}(d) and $\eta$ is the detection efficiency at each detector.  The first term is due to the heralded three-photon N00N state term, i.e., the first term in eq.~(\ref{state}), and the second/third terms are from the second term in eq.~(\ref{state}). The amplitudes expressed as $|\textrm{etc}\rangle$ in eq.~(\ref{state}) do not contribute to the outcome of the three-photon detector. The experimental data in Fig.~\ref{temporal}(b) is in good agreement with the above theoretical calculation. In Fig.~\ref{temporal}(c), we show the heralded three-photon N00N state interference observed in four-fold coincidences. The data, however, is affected by the accidental coincidences due to non-N00N state terms, i.e., 2nd and 3rd terms in eq.~(\ref{prob}). (As mentioned before, the non-N00N state contributions can be removed by a trigger-operated shutter in mode 2.) When the accidental contribution is subtracted, however, the four-fold coincidence shown in Fig.~\ref{temporal}(d) exhibits high-visibility ($V=0.86 \pm 0.08$) heralded three-photon N00N state interference with three-times greater phase resolution than the single-photon case shown in Fig.~\ref{temporal}(a). (Only the first term in eq.~(\ref{prob}) is now relevant.) Even at much higher pump power of 300 mW, similar results are observed, see Fig.~\ref{temporal}(e) and Fig.~\ref{temporal}(f), albeit with somewhat reduced visibility ($V= 0.74 \pm 0.07$). Potential sources for the visibility reduction at a higher pump power are mode mismatch, and increased double-pair and triple-pair accidentals. The observed three-photon N00N state visibilities, however, are well above the classical limit of 0.1 \cite{bentley04,afek10prl}.


Having confirmed that the three-photon polarization N00N state, $|\psi\rangle_p$, is prepared with sufficiently high purity, we now proceed to demonstrate the Young's double-slit interference with the N00N state $|\psi\rangle_s$. For this measurement, QWP2 is now restored at its original location and the $|\psi\rangle_p$ is transformed to the $|\psi\rangle_s$ state with the mode converter shown in Fig.~\ref{setup}(b). Spatial interference fringes are measured with the detection scheme shown in Fig.~\ref{setup}(d) and the pump power is increased to 400 mW.

\begin{figure}[t]
   \includegraphics[width=3in]{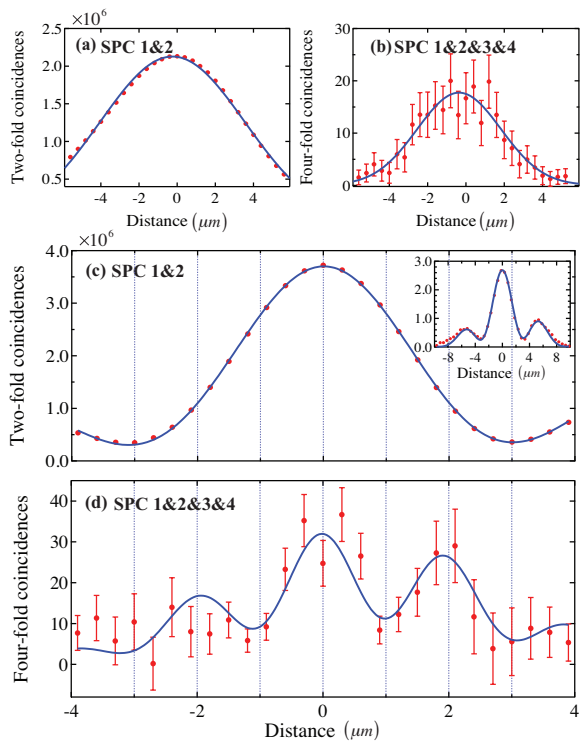}
  \caption{Spatial profile of the heralded (a) single-photon state and (b) three-photon state. The data accumulation time is 3400 s each point and solid lines are Gaussian fits to the data.  Interference of the heralded single-photon state and the heralded three-photon N00N state are shown in (c) and in (d), respectively. The data were accumulated for 1700 s each point. Solid lines are fitting curves based on theoretical calculation.  Accidental coincidences have been subtracted. 
The $|\psi\rangle_s$ state  exhibits three times faster spatial interference fringes (2.0 $\mu$m) than that of the single-photon state (6.0 $\mu$m).
     }
  \label{spatial}
  \end{figure}

The experimental data for the double-slit interference of the three-photon N00N state $|\psi\rangle_s$ are shown in Fig.~\ref{spatial}. We first measured the spatial profile of the beam at the focus of the lens by blocking mode $b$. The single-photon and the three-photon spatial profiles are shown in Fig.~\ref{spatial}(a) and Fig.~\ref{spatial}(b), respectively. Note that both the single-photon and the three-photon states are heralded by single-photon detection at SPC1. Fitting the data with a Gaussian function $\exp[-(x/w_0)^2]$, we find that the $2 w_0$ widths are $10.8 \pm 0.4$ $\mu$m for the single-photon case and $6.2 \pm 0.6$ $\mu$m for the three-photon case. This is in good agreement with the theoretical estimation that the three-photon probability distribution is proportional to the cube of the single-photon one.

For the Young-type double-slit interference measurement, both modes $a$ and $b$ are now open and interference fringes are measured in two-fold and four-fold coincidences at the focus of the lens as a function of the scanning fiber tip position, see Fig.~\ref{spatial}(c) and Fig.~\ref{spatial}(d). The spatial interference fringe spacing is determined by the beam spacing $d$ and the focal length of the lens (15 mm). In theory, the fringe patterns are expected to show sinusoidal modulations within the respective spatial profiles of the single- and three-photon states and the modulation frequency for the three-photon N00N state should be three times more than that of the single-photon state. The experimental data, which show fringe spacing of 6.0 $\mu$m  for the single-photon state and $2.0$ $\mu$m for the three-photon N00N state, are thus in good agreement with the theory. We point out that, although the mode field diameter (5.6 $\mu$m) of the scanning fiber tip is larger than the fringe spacing due to the three-photon N00N state, the fact that the fiber is single-mode at 780 nm allows us to measure a spatial fringe spacing much smaller than the mode field diameter \cite{yariv, jlee}.

In general, spatial super-resolution itself may not necessarily be of quantum origin \cite{yab}. However, the fringe visibility $V$ is an important feature that distinguishes between quantum and classical cases. For three-photon Young's double-slit interference, the classical limit of the fringe visibility  can be calculated by considering intensity-cubed detection (i.e., three-photon detection) rather than detection linear in intensity (i.e., single-photon detection) and has been shown to be 0.1 \cite{bentley04,afek10prl}. In this work, the double-slit interference with the three-photon N00N state exhibits $V= 0.49\pm0.09$ which is well above the classical limit of 0.1.

These experimental results are the first demonstration of the Young-type double-slit interference of the three-photon N00N state. Our work demonstrates clearly both spatial super-resolution and fringe visibility surpassing the classical limit. This is the first time that the spatial quantum interference of the N00N state is observed for more than two photons, thus paving the way towards new applications in quantum metrology, quantum imaging, and quantum interferometric lithography.



This work was supported by the National Research Foundation of Korea (2009-0070668 and 2009-0084473) and  the KRISS Single-Quantum-Based Metrology in Nanoscale Project. YSK acknowledges the support of the Korea Research Foundation (KRF-2007-511-C00004).




\begin{thebibliography}{99}



\bibitem{feynman} R.P. Feynman, R.B. Leighton, and M. Sands, \textit{The Feynman Lectures on Physics} (Addison-Wesley, Reading, 1965), Vol. III.

\bibitem{kim00} Y.-H. Kim \textit{et al.}, Phys. Rev. Lett. \textbf{81}, 1 (2000).


\bibitem{sinha10} U. Sinha \textit{et al.}, Science \textbf{329}, 418 (2010).


\bibitem{jacobson95} J. Jacobson \textit{et al.}, Phys. Rev. Lett. \textbf{74}, 4835 (1995).


\bibitem{ourj} A. Ourjoumtsev \textit{et al.}, Nature \textbf{448}, 784 (2007).

\bibitem{dow} J. Dowling, Contemp. Phys. \textbf{49}, 125 (2008).

\bibitem{boto00} A.N. Boto \textit{et al.}, Phys. Rev. Lett.  \textbf{85}, 2733 (2000).

\bibitem{bjork01} G. Bj$\rm{\ddot o}$rk, L.L. Sanchez-Soto, and J. Soderholm, Phys. Rev. Lett. \textbf{86}, 4516 (2001).

\bibitem{giovannetti04} V. Giovannetti, S. Lloyd, and L. Maccone, Science \textbf{306}, 1330 (2004).

\bibitem{kwon10}  O. Kwon, Y.-S. Ra, and Y.-H. Kim, Phys. Rev. A  \textbf{81}, 063801 (2010).


\bibitem{fiu} J. Fiurasek, Phys. Rev. A \textbf{65}, 053818 (2002).

\bibitem{cable07} H. Cable, and J. P. Dowling, Phys. Rev. Lett. \textbf{99}, 163604 (2007).

\bibitem{kapale07} K.T. Kapale and J.P. Dowling, Phys. Rev. Lett. \textbf{99}, 053602 (2007). 

\bibitem{dan} M. D'Angelo, A. Garuccio, and V. Tamma, Phys. Rev. A \textbf{77}, 063826 (2008).






\bibitem{eda02} K. Edamatsu, R. Shimizu, and T. Itoh, Phys. Rev. Lett. \textbf{89}, 213601 (2002).

\bibitem{dangelo01} M. D'Angelo, M.V. Chekhova, and Y.H. Shih, Phys. Rev. Lett. \textbf{87}, 013602 (2001).

\bibitem{kawabe07} Y. Kawabe \textit{et al.}, Opt. Express \textbf{15}, 14244 (2007).


\bibitem{mitchell04} M.W. Mitchell, J.S. Lundeen, and A.M. Steinberg, Nature \textbf{429}, 161 (2004).

\bibitem{kim09} H. Kim, H.-S. Park, and S.-K. Choi, Opt. Express \textbf{17}, 19720 (2009).

\bibitem{walther04} P. Walther \textit{et al.}, Nature \textbf{429}, 158 (2004).

\bibitem{nagata07} T. Nagata \textit{et al.}, Science \textbf{316}, 726 (2007).

\bibitem{sun06} F. W. Sun \textit{et al.}, Phys. Rev. A \textbf{74}, 033812 (2006).

\bibitem{afek10} I. Afek, O. Ambar, and Y. Silberberg, Science \textbf{328}, 879 (2010).


\bibitem{hom} C.K. Hong, Z.Y. Ou, and L. Mandel, Phys. Rev. Lett. \textbf{59}, 2044 (1987).


\bibitem{bentley04} S.J. Bentley, and R.W. Boyd, Opt. Express \textbf{12}, 5735 (2004).
\bibitem{afek10prl} I. Afek, O. Ambar, and Y. Silberberg, Phys. Rev. Lett. \textbf{104}, 123602 (2010).


\bibitem{yariv} A. Yariv, and P. Yeh, \textit{Photonics: Optical Electronics in Modern Communications}, 6th ed., (Oxford University Press, USA, 2006) 



\bibitem{jlee} J.-C. Lee, H.-S. Park and Y.-H. Kim, unpublished.

\bibitem{yab} E. Yablonovitch and R.B. Vrijen, Opt. Eng. \textbf{38}, 334 (1999).










\end{thebibliography}
\end{document}